\newcommand*{\ATLASLATEXPATH}{latex/}
\documentclass[UKenglish,texlive=2011,cernpreprint,txfonts]{\ATLASLATEXPATH atlasdoc}

\usepackage{\ATLASLATEXPATH atlaspackage}
\usepackage{\ATLASLATEXPATH atlasbiblatex}
\usepackage{\ATLASLATEXPATH atlascontribute}
\usepackage{\ATLASLATEXPATH atlasphysics}

\graphicspath{{logos/}{figures/}}

\usepackage{VHPaper-defs}
\usepackage{graphicx}
\usepackage{epstopdf}

\AtlasTitle{Search for new resonances decaying to a $W$ or $Z$ boson and a
Higgs boson in the $\ell^+ \ell^- b\bar b$, $\ell \nu b\bar b$, and $\nu\bar{\nu}
b\bar b$ channels with $pp$ collisions at $\sqrt s = 13$~TeV with the
ATLAS detector}

\author{The ATLAS Collaboration}

\date{\today}

\PreprintIdNumber{CERN-EP-2016-148}

\AtlasJournal{Phys.\ Lett.\ B.}
\AtlasAbstract{%
A search is presented for new resonances decaying to a $W$ or $Z$ boson
and a Higgs boson in the $\ell^+ \ell^- b\bar b$, $\ell\nu b\bar b$, and
$\nu\bar{\nu} b\bar b$ channels in $pp$ collisions at $\sqrt s = 13$~TeV
with the ATLAS detector at the Large Hadron Collider using a total
integrated luminosity of $3.2$ fb$^{-1}$.
The search is conducted by looking for a localized excess in the
$WH$/$ZH$ invariant or transverse mass distribution.
No significant excess is observed, and the results are interpreted
in terms of constraints on a simplified model based on a
phenomenological Lagrangian of heavy vector triplets.
}

\hypersetup{pdftitle={ATLAS draft},pdfauthor={The ATLAS Collaboration}}

\begin{document}

\maketitle

%-------------------------------------------------------------------------------
\section{Introduction}
\label{sec:intro}
The Higgs boson discovery by the ATLAS~\cite{:2012gk} and
CMS~\cite{:2012gu} collaborations imposes constraints on theories
beyond the Standard Model (SM). Nevertheless, quadratically divergent
radiative corrections to the Higgs boson mass make it unnatural for
the SM to be valid beyond a scale of a few TeV.
Various dynamical electroweak symmetry-breaking scenarios attempt to
solve the naturalness problem by assuming a new strong interaction at
a higher scale. These models generically predict the existence of new
resonances decaying to a vector boson plus the Higgs boson, as for
example in Minimal Walking
Technicolour~\cite{Sannino:2004qp,Foadi:2007ue,Belyaev:2008yj},
Little Higgs~\cite{Schmaltz:2005ky}, or composite Higgs
models~\cite{Dugan:1984hq,Agashe:2004rs}.

This Letter describes a search for new heavy vector bosons decaying to a SM
vector boson and a SM Higgs boson, denoted hereafter by ${W}^\prime$
and ${Z}^\prime$ ($pp \rightarrow {W}^\prime \rightarrow {WH}$ and
$pp \rightarrow {Z}^\prime \rightarrow {ZH}$) and together as $V^\prime$.
The analyses described here only consider leptonic decays of the vector
bosons (${W}\rightarrow \ell \nu$, ${Z}\rightarrow \ell^+ \ell^-$,
${Z}\rightarrow\nu\bar{\nu}$; $\ell = e,\mu$) and decays of the Higgs boson to
bottom-quark pairs (${H}\rightarrow b\bar{b}$).
This results in three search channels: ${ W}^\prime \rightarrow { W H}
\rightarrow \ell \nu b\bar{b}$, ${Z}^\prime \rightarrow {Z H} \rightarrow
\ell^+ \ell^- b\bar{b}$, and ${ Z}^\prime \rightarrow { Z H} \rightarrow \nu \bar{\nu}
b\bar{b}$.  

For the interpretation of the results in terms of a search for heavy vector bosons,
a simplified model~\cite{Pappadopulo:2014qza} based on a phenomenological
Lagrangian is used as a benchmark. This simplified model incorporates
heavy vector triplets (HVT) and allows for the interpretation of
the results in a large class of models.
Here, the new heavy vector bosons couple to the Higgs boson and SM
gauge bosons via a combination of parameters $g_Vc_H$ and to the
fermions via the combination $(g^2/g_V)c_F$, where $g$ is the weak
SU(2) coupling constant.
The parameter $g_V$ represents the strength of the new vector boson's
interaction, and $c_H$ and $c_F$ are multiplicative factors to modify
the couplings to the Higgs boson and the fermions, and are expected to
be of order unity in most models.
Two benchmark models~\cite{Pappadopulo:2014qza} are used here. 
In the first, referred to  as \textit{Model A}, the branching
fractions to fermions and gauge bosons are comparable, as in some
extensions of the SM gauge group~\cite{Barger:1980ix}.
For \textit{Model B}, fermionic couplings are suppressed, for example
in a composite Higgs model~\cite{Contino:2011np}.
The regions of HVT parameter space probed in this Letter correspond to
the production of resonances with an intrinsic width that is narrow
relative to the experimental resolution, which is roughly 10\% of the
resonance mass.

Previous searches in the same final states have been performed by both
the ATLAS and CMS collaborations using data at $\sqrt s = 8$~TeV.
The ATLAS searches for ${V}^\prime \rightarrow {VH}$ set a lower
limit at the 95\% confidence level (CL) on the ${W}^\prime$
(${Z}^\prime$) mass at 1.47~(1.36)~TeV, assuming the HVT benchmark
\textit{Model A} with $g_V=1$~\cite{Aad:2015yza}.
Searches by the CMS Collaboration for ${V}^\prime \rightarrow {VH}$,
based on HVT benchmark \textit{Model B} with $g_V=3$, similarly
exclude heavy resonance masses up to 1.1~TeV (${Z}^\prime \rightarrow
{ZH}$), 1.5~TeV (${W}^\prime \rightarrow {WH}$), yielding a combined
limit of 1.7~TeV (${V}^\prime \rightarrow {VH}$) in the fully hadronic
final state~\cite{Khachatryan:2015bma}, and masses up to 1.5~TeV for
the ${W}^\prime \rightarrow {WH} \rightarrow \ell \nu b \bar b$ final
state~\cite{Khachatryan:2016yji}.
A search by the CMS Collaboration has been carried out for a narrow
resonance decaying to $ZH$ in the $\tau^+ \tau^- b \bar b$ final
state, setting limits on the production cross-section of $Z^{\prime}$
assuming the HVT benchmark \textit{Model B} with
$g_V=3$~\cite{Khachatryan:2015ywa}.
The ATLAS Collaboration has also performed a search for narrow
resonances decaying to $VV$ final states~\cite{Aad:2016ipg}.

The search presented here has been optimized to be sensitive to
resonances of mass larger than 1 TeV, hence decaying to highly boosted
final-state particles.
As a consequence, the Higgs boson decay to bottom quarks is less
likely to be observed as two separate jets than as a single wide jet
where the two $b$-jets are ``merged'' (the Higgs boson candidate).
Bottom-quark tagging is used as a means to further purify the event
selection. Decays of the Higgs boson to charm quarks are included in
the signal Monte Carlo simulation to properly account for the small
contribution of $b$-tagged charm quarks.
Together, the reconstructed mass of the Higgs boson candidate jet and
the results of the bottom-quark tagging are used to identify likely
Higgs boson candidates.
The search is performed by examining the distribution of the
reconstructed ${VH}$ mass  ($m_{VH}$)  or transverse mass
($m_{{\rm T}, VH}$) for a localized excess.
The signal strength and background normalization are determined
from a binned maximum-likelihood fit to the data distribution in each
channel and are used to evaluate bounds on the production
cross-section times decay branching fraction for $V^\prime$ bosons.

%-------------------------------------------------------------------------------

%-------------------------------------------------------------------------------
\section{ATLAS detector}
\label{sec:detector}
%-------------------------------------------------------------------------------

The ATLAS detector~\cite{Aad:2008zzm} is a general-purpose particle detector used to
investigate a broad range of physics processes.
It includes inner tracking devices surrounded by
a superconducting solenoid, electromagnetic and hadronic calorimeters and a muon spectrometer
with a toroidal magnetic field. 
The inner detector consists of a high-granularity silicon pixel
detector, including the insertable B-layer~\cite{ibl} installed after
Run 1 of the LHC, a silicon strip detector, and a straw-tube tracker;
it is situated inside a 2 T axial field and provides precision
tracking of charged particles with pseudorapidity $|\eta| < 2.5$,
where the pseudorapidity is defined in terms of the
polar angle\footnote{ATLAS uses a right-handed coordinate system with its
origin at the nominal interaction point (IP) in the centre of the
detector and the $z$-axis along the beam axis.
The $x$-axis points from the IP to the centre of the LHC ring, and the
$y$-axis points upward.
Cylindrical coordinates $(r,\phi)$ are used in the transverse plane,
$\phi$ being the azimuthal angle around the $z$-axis.} $\theta$ as
$\eta=-\ln\tan(\theta/2)$.
The straw-tube tracker also provides transition radiation measurements
for electron identification up to $|\eta| =  2.0$.
The calorimeter system covers the pseudorapidity range $|\eta| < 4.9$.
It is composed of sampling calorimeters with either liquid argon
or scintillator tiles as the active media. The muon spectrometer
provides muon identification and measurement for $|\eta| < 2.7$.
The ATLAS detector has a two-level trigger system to select events for
offline analysis~\cite{ATL-DAQ-PUB-2016-001}.

%-------------------------------------------------------------------------------
\section{Data and simulated samples}
\label{sec:samples}
%-------------------------------------------------------------------------------
The data used in this analysis were recorded with the ATLAS detector
during the 2015 $pp$ collisions run and correspond to a total
integrated luminosity of $3.2$ fb$^{-1}$~\cite{ATLAS_luminosity} at
$\sqrt{s}=$ 13 TeV.
Collision events satisfy a number of requirements ensuring that the
ATLAS detector was operating in stable conditions while the data were
recorded.

Simulated Monte Carlo (MC) samples for the HVT are generated with
MadGraph5\_aMC@NLO 2.2.2~\cite{Alwall:2014hca} using the {\sc
NNPDF2.3LO}~\cite{Ball:2011mu} parton distribution functions  (PDFs).
For all signal events, parton showering and hadronization are
performed with {\sc Pythia} 8.186~\cite{Pythia8} using the A14 set of
tuned parameters (tune)~\cite{ATL-PHYS-PUB-2014-021}.
The Higgs boson has its mass set to 125.5 GeV, and it is allowed to
decay to $b\bar{b}$ and $c\bar{c}$ pairs, with relative branching
fractions BR$(H\rightarrow c\bar{c})/$BR$(H\rightarrow b\bar{b})=0.05$
fixed to the Standard Model prediction~\cite{Heinemeyer:2013tqa}.
Signal samples are generated for a range of resonance masses from
$0.7$ to $5\,\TeV$ in steps of 100 GeV up to 2 TeV and in wider steps
for higher masses.

Monte Carlo samples are used to model the shape and normalization of
most SM background processes.
Diboson events ($WW$, $WZ$, $ZZ$) and events containing a $W$ or $Z$
boson with associated jets ($W$+jets, $Z$+jets) are simulated using
the {\sc Sherpa} 2.1.1~\cite{SHERPA} generator.
Matrix elements are calculated using the {\sc
Comix}~\cite{Gleisberg:2008fv} and {\sc
OpenLoops}~\cite{Cascioli:2011va} matrix element generators and merged
with the {\sc Sherpa} parton shower using the {\sc ME+PS@NLO}
prescription~\cite{Hoeche:2012yf}.
For $W$+jets and $Z$+jets events these are calculated for up to two
additional partons at next-to-leading order (NLO) and four partons at
leading order (LO); they are calculated
for up to one ($ZZ$) or no ($WW$, $WZ$) additional partons at NLO and
up to three additional partons at LO.
The CT10 PDF set~\cite{Lai:2010vv} is used in conjunction with
dedicated parton shower tuning developed by the authors of {\sc
Sherpa}.

The $W/Z$+jets simulated samples are split into different components
according to the true flavour of the jets, i.e. $W/Z+q$,  where $q$
denotes a light quark ($u$, $d$, $s$) or a gluon, $W/Z+c$  and
$W/Z+b$.
Each event is categorized based on the hadrons associated to the track
jets matched to the highest-$p_{\rm T}$ (leading) large-$R$ jet in the
event; these jet collections are introduced in Section~\ref{sec:object}.
If there is an associated bottom (charm) hadron, then the event is
given a $b$ ($c$) label; otherwise it is labelled $W/Z+q$.

For the generation of $t\bar t$ and single top quarks in the $Wt$- and
s-channels the {\sc Powheg-BOX} v2~\cite{Powheg1,Powheg2,Powheg3}
generator with the {\sc CT10} PDF sets is used. Electroweak t-channel
single-top-quark events are generated using the {\sc Powheg-BOX} v1
generator. This generator uses the four-flavour scheme for the NLO
matrix elements calculations together with the four-flavour PDF set
\cite{Lai:2010vv}. For all top processes, top-quark spin correlations
are preserved (for the t-channel, top quarks are decayed using {\sc
MadSpin}~\cite{Artoisenet:2012st}). The parton shower, fragmentation,
and the underlying event are simulated using {\sc
Pythia 6.428}~\cite{Sjostrand:2006za,} with the {\sc
CTEQ6L1}~\cite{Pumplin:2002vw} PDF sets and the corresponding Perugia
2012 tune (P2012)~\cite{Skands:2010ak}. The top quark mass is set to
172.5 GeV.
The {\sc EvtGen} v1.2.0 program~\cite{Lange:2001uf} is used for the
bottom and charm hadron decays.

Finally, SM Higgs boson production in association with a $W/Z$ boson
is simulated using {\sc Pythia 8.186} and {\sc Powheg} with showering
by {\sc Pythia 8.186} for the gluon-induced associated production; the
{\sc CT10} PDFs and the AZNLO tune is used in both
cases~\cite{AZNLO:2014}.
SM Higgs boson production is considered as a background in this
search.

All simulated MC samples include the effect of multiple $pp$
interactions in the same and neighbouring bunch crossings (pile-up) by
overlaying simulated minimum-bias events on each generated signal or
background event.
Simulated events are reconstructed with the standard ATLAS
reconstruction software used for collision data.

%-------------------------------------------------------------------------------
\section{Object selection}
\label{sec:object}
%-------------------------------------------------------------------------------
\def\mptvec{\ensuremath{\vec{p}_{\mathrm{T}}^{\mathrm{~miss}}}}
\def\metvec{\ensuremath{\vec{E}_{\mathrm{T}}^{\mathrm{miss}}}}
\def\mpt{\ensuremath{p_\mathrm{T}^{\mathrm{miss}}}}

Collision vertices are reconstructed from tracks with transverse
momentum $\pt >$ 400 MeV.
If an event contains more than one vertex candidate, the one with the
highest $\sum\pt^{2}$ calculated considering all the associated tracks
is selected as the primary vertex.

Electrons are reconstructed from inner-detector tracks that are
matched to energy clusters in the electromagnetic calorimeter
obtained using the standard ATLAS sliding-window
algorithm~\cite{Aad:2014nim}.
Electron candidates satisfy criteria for the electromagnetic shower
shape, track quality and track-cluster matching.
These requirements are applied using a likelihood-based approach, and
two different working points are used: ``loose'' and ``tight'' with
increasing purity~\cite{Aad:2014fxa}.
Muons are identified by matching tracks found in the inner detector to
either full tracks or track segments reconstructed in the muon
spectrometer~\cite{Aad:2016jkr}.
Muons are required to pass identification requirements based on
quality criteria imposed on the inner detector and muon spectrometer
tracks, and, as for electrons, both ``loose'' and ``tight'' operating
points are used.
Both the electrons and muons are required to have a minimum  \pt\ of
$7$ GeV and to lie within a region with a good reconstruction and
identification efficiency ($|\eta| < 2.7$ for muons and $|\eta|<2.47$
for electrons).
They are required to be isolated using requirements on the sum of the
\pt\ of the tracks lying in a cone around the lepton direction whose
radius, $\Delta R = \sqrt{(\Delta \eta)^2 + (\Delta \phi)^2}$,
decreases as a function of the lepton \pt .
They are also required to originate from the primary
vertex~\cite{Aad:2014fxa,Aad:2016jkr}.

Three types of jets are used to characterize the hadronic activity of
events: large-$R$ jets, small-$R$ jets  and track jets.
All three jet collections are reconstructed using the anti-$k_t$
algorithm but with different radius parameters, $R$~\cite{Cacciari:2008gp} .
Large- and small-$R$ jets are built from noise-suppressed topological
clusters~\cite{Aad:2016upy} in the calorimeter, while track jets are
constructed from inner-detector tracks.

Large-$R$ jets are constructed with a radius parameter $R = 1.0$.
They are required to have $\pT>250$ GeV and $|\eta|<2.0$.
These jets are trimmed~\cite{Krohn:2009th} to suppress the energy of
clusters which originate from initial-state radiation, pile-up
vertices or the underlying event.
This is done by reclustering the constituents of the initial jet
using the \kt\ algorithm~\cite{Catani:1993hr} into subjets of radius $R_{{\rm
sub}}$; the constituents of any subjet with transverse momentum
less than $f_{{\rm cut}}$ times the transverse momentum of the
initial jet are removed.
The $R_{{\rm sub}}$ and $f_{{\rm cut}}$ parameter values found to be
optimal in identifying hadronic $W/Z$ boson
decays~\cite{ATL-PHYS-PUB-2015-033} are $R_{{\rm sub}} = 0.2$ and
$f_{{\rm cut}} = 5\%$.
Large-$R$ jets are required to be separated by $\Delta R > 1.0$ to
the nearest electron candidate.

Small-$R$ jets are reconstructed with a radius parameter \mbox{$R =
0.4$} and are required to have $\pT>20$~GeV and $|\eta|<2.4$ or $\pt >
30$~GeV and $2.4 < |\eta| < 4.5$.
If an electron candidate has an angular separation $\Delta R < 0.2$ to
a small-$R$ jet, the small-$R$ jet is discarded; however, if an
electron candidate and small-$R$ jet are separated by $0.2 < \Delta R
< 0.4$, the electron candidate is removed.
Similarly, if a small-$R$ jet is separated by $\Delta R < 0.4$ to the
nearest muon candidate, the small-$R$ jet is discarded if it has fewer
than three associated inner-detector tracks; otherwise the muon
candidate is removed.
The jet-vertex-tagger discriminant is used to reject small-$R$ jets
originating from pile-up based on vertex information of each of the
jet's associated tracks~\cite{Aad:2015ina}.
Small-$R$ jets with $\pt < 50$~GeV and $|\eta| < 2.4$ must have a
discriminant greater than 0.64.
The energies of both the large-$R$ and small-$R$ jets and the mass of
the large-$R$ jets are corrected for energy losses in passive
material, for the non-compensating response of the calorimeter, and
for any additional energy due to multiple $pp$
interactions~\cite{Aad:2011he}.

The third type of jet used in this analysis, track jets, are built
from inner-detector tracks  with \pt\ $>$ 400 MeV associated with the
primary vertex
using the anti-\kt\ algorithm with $R = 0.2$ and are required to
have \pt\ $> 10$ GeV and $|\eta| < 2.5$. Track jets containing
$b$-hadrons are identified using the MV2c20 $b$-tagging
algorithm~\cite{Aad:2015ydr,ATL-PHYS-PUB-2015-022} with $70\%$
efficiency and a rejection factor of about 5.6 (180) for jets
containing $c$-hadrons (not containing $b$- or $c$-hadrons) in a
simulated sample of $t\bar{t}$ events and are matched to the
large-$R$ jets via ghost-association~\cite{Cacciari:2008gp}.

Hadronically decaying $\tau$-lepton candidates, which are used to veto
background events, are reconstructed from noise-suppressed topological
clusters in the calorimeter using the anti-\kt\ algorithm with $R =
0.4$.
They are required to have $\pT>20$ GeV, $|\eta| < 2.5$ and to be
outside the transition region between the barrel and end-cap
calorimeters ($1.37 < |\eta| < 1.52$); to have
either one or three associated tracks; and to satisfy the ``medium”
working point criteria~\cite{ATL-PHYS-PUB-2015-045}.
The leptonic decays of $\tau$ leptons are simulated and included in
the acceptance if the final-state electron or muon passes lepton
selections.

The presence of one or more neutrinos in collision events can be
inferred from an observed momentum imbalance in the transverse plane.
The missing transverse momentum ($\metvec$) is calculated as the
negative vectorial sum of the transverse momenta of all the muons,
electrons, small-$R$ jets, and any inner-detector tracks from the
primary vertex not matched to any of these
objects~\cite{ATL-PHYS-PUB-2015-023}.
The magnitude of the $\metvec$ is denoted by $\met$.
For multi-jet background rejection, a similar quantity, $\mptvec$, is
computed using only charged-particle tracks originating from the
nominal hard-scatter vertex, and its magnitude is denoted by $\mpt$.

%-------------------------------------------------------------------------------
\section{Event selection}
\label{sec:selection}
%-------------------------------------------------------------------------------
\def\met{\ensuremath{E_{\mathrm{T}}^{\mathrm{miss}}}}
\def\mht{\ensuremath{H_{\mathrm{T}}^{\mathrm{miss}}}}

This analysis is performed for events containing zero, one, or two
charged leptons (electrons or muons), targeting the $Z^\prime
\rightarrow Z H \rightarrow \nu \bar{\nu} b\bar{b}$ , $W^\prime
\rightarrow W H \rightarrow \ell \nu b\bar{b}$  and $Z^\prime
\rightarrow  Z H\rightarrow \ell^+ \ell^- b\bar{b}$  decay modes,
respectively; the ``loose'' lepton identification working points are
used to categorize events by their charged-lepton number.
While the 1-lepton channel has some acceptance for the $Z^\prime
\rightarrow Z H \rightarrow \ell \ell b\bar{b}$ signal, it has
significantly larger backgrounds than the 2-lepton channel; the
1-lepton channel is therefore not included in the $Z^\prime$ search.
The 0-lepton channel has a non-negligible acceptance for the
$W^\prime \rightarrow W H \rightarrow \ell \nu b\bar{b}$ signal in
events in which the lepton is not detected or is a hadronically
decaying $\tau$-lepton; it also has smaller predicted backgrounds than
the 1-lepton channel.
For this reason, the 0-lepton channel and the 1-lepton channel are
combined in the $W^\prime$ search.

In the 0-lepton channel events are recorded using an \met\ trigger
with an online threshold of 70 GeV, while in the 2-lepton channel,
events are recorded using a combination of single-lepton triggers,
with the lowest \pt\  threshold being 24 GeV for isolated electrons
and 20 GeV for isolated muons. These triggers are complemented with
non-isolated ones with higher \pt\ thresholds. The 1-lepton channel
uses the single-electron triggers for the electron channel and a
combination of the \met\ trigger and single-muon trigger for the muon
channel, where the \met\ trigger considers only the energy of objects
in the calorimeter, and thus muons are seen as a source of \met.
For events selected by lepton triggers, the object that satisfied the
trigger is required to be matched geometrically to the
offline-reconstructed lepton.

Events containing no loose lepton are assigned to the 0-lepton channel.
The multi-jet and non-collision backgrounds in the 0-lepton channel
are suppressed by imposing requirements on \mpt\ (\mpt\ $>30$~GeV),
\met\ (\met\ $>200$~GeV), the azimuthal angle between \metvec\ and
\mptvec\ ($\Delta \phi(\metvec, \mptvec)<\pi/2$), and the azimuthal
angle between \metvec\ and the leading large-$R$ jet ($\Delta \phi
(\metvec,$~large-$R~{\rm jet})>2\pi/3$).
An additional requirement is imposed on the azimuthal angle between
\metvec\ and the nearest small-$R$ jet that is not identified as a
$\tau$-lepton
(${\rm min}[\Delta\phi(\metvec,$~small-$R~{\rm jet})]>\pi/9$). 
Finally, only in the search for $Z^\prime \rightarrow ZH$, events
containing one or more identified hadronically decaying $\tau$-lepton
candidates are rejected; this veto reduces the $W+$jets and
$t\bar{t}$ contributions and has a negligible impact on the
$Z^\prime$ acceptance.
Since it is not possible to fully reconstruct the invariant mass of
the $ZH\rightarrow \nu\nu b\bar{b}$ system due to the neutrinos
present in the final state, the transverse mass is used as the final
discriminant: \( m_{{\rm T}, VH}= \sqrt{(E_{{\rm T}}^{\rm jet} +
{\met})^2 - (\vec{p}_{\rm T}^{\mathrm{~jet}} + \metvec)^2} \), where
$\ensuremath{{{\pt^{\rm jet}}}}$ ($\ensuremath{{{E_{{\rm T}}^{\rm
jet}}}}$) is the transverse momentum (energy) of the leading large-$R$
jet.

Events containing exactly one lepton with $\pt > 25$~GeV (and with $ |\eta| < $2.5 for muons)  are 
assigned to the 1-lepton channel.
To reduce the multi-jet background from non-prompt leptons or from
jets faking leptons, the lepton must satisfy the tight quality
criteria.
Moreover, isolation requirements based on the sum of the calorimeter
energy deposits and track transverse momentum in a cone around the
lepton direction are applied.
In addition, the event must have significant missing transverse
momentum: $\met> 100$ GeV.
To reconstruct the invariant mass of the $WH\rightarrow  \ell \nu
b\bar{b}$ system in the 1-lepton channel, the momentum of the neutrino
in the $z$-direction, $p_{z}$, is obtained by imposing the $W$ boson
mass constraint on the lepton--neutrino system.
In the resulting quadratic equation, $p_{z}$ is taken as either the
real component in the case of complex solutions or the smaller of the two
solutions if both solutions are real.

Events containing exactly two loose leptons of the same flavour with
$\pt > 25$~GeV (and with $ |\eta| < $2.5 for muons) are assigned to the 2-lepton channel.
Only loose track isolation requirements are applied since this channel has negligible background from fake and non-prompt leptons.
The invariant mass of the two leptons, $m_{\ell\ell}$, must be in the range 70--110~GeV for the dielectron selection.
This range is widened to 55--125 GeV for the dimuon selection due to
the poorer momentum  resolution at high \pt.
To improve the $m_{\rm VH}$ resolution of $ZH\rightarrow \mu \mu
b\bar{b}$ events, the four-momentum of the dimuon system is scaled by
$m_{Z }/m_{\mu\mu}$, where $m_{Z }=91.2$~GeV and $m_{\mu\mu}$ is the
invariant mass of the dimuon system.

All three channels require  at least one large-$R$ jet with $\pt >
250$~GeV and $|\eta| < 2.0$.
The leading large-$R$ jet is considered to be the $H\rightarrow
b\bar{b}$ candidate.
To enhance the sensitivity to a $VH$ signal, the leading large-$R$
jet is required to have at least one associated track jet, and at
least one of the associated track jets must be
$b$-tagged~\cite{ATL-PHYS-PUB-2015-035}.
If more than two track jets are matched to the $H\rightarrow
b\bar{b}$ candidate, only the two with the highest $\pt$ are
considered for the $b$-tagging requirement.
In all the three channels, events are vetoed if they have at least
one $b$-tagged track jet not matched to the leading large-$R$ jet.
This veto is particularly effective in suppressing the $t\bar{t}$
background in the 0- and 1-lepton channels.
The events fulfilling these requirements are divided into 1- and 2
$b$-tag categories depending on whether one or both of the two leading
track jets matched to the leading large-$R$ jet are $b$-tagged.

\begin{figure}[tb]
\centering
\subfigure[][0-lepton, $\geq 1$ $b$-tag]{
  \includegraphics[width=0.45\textwidth]{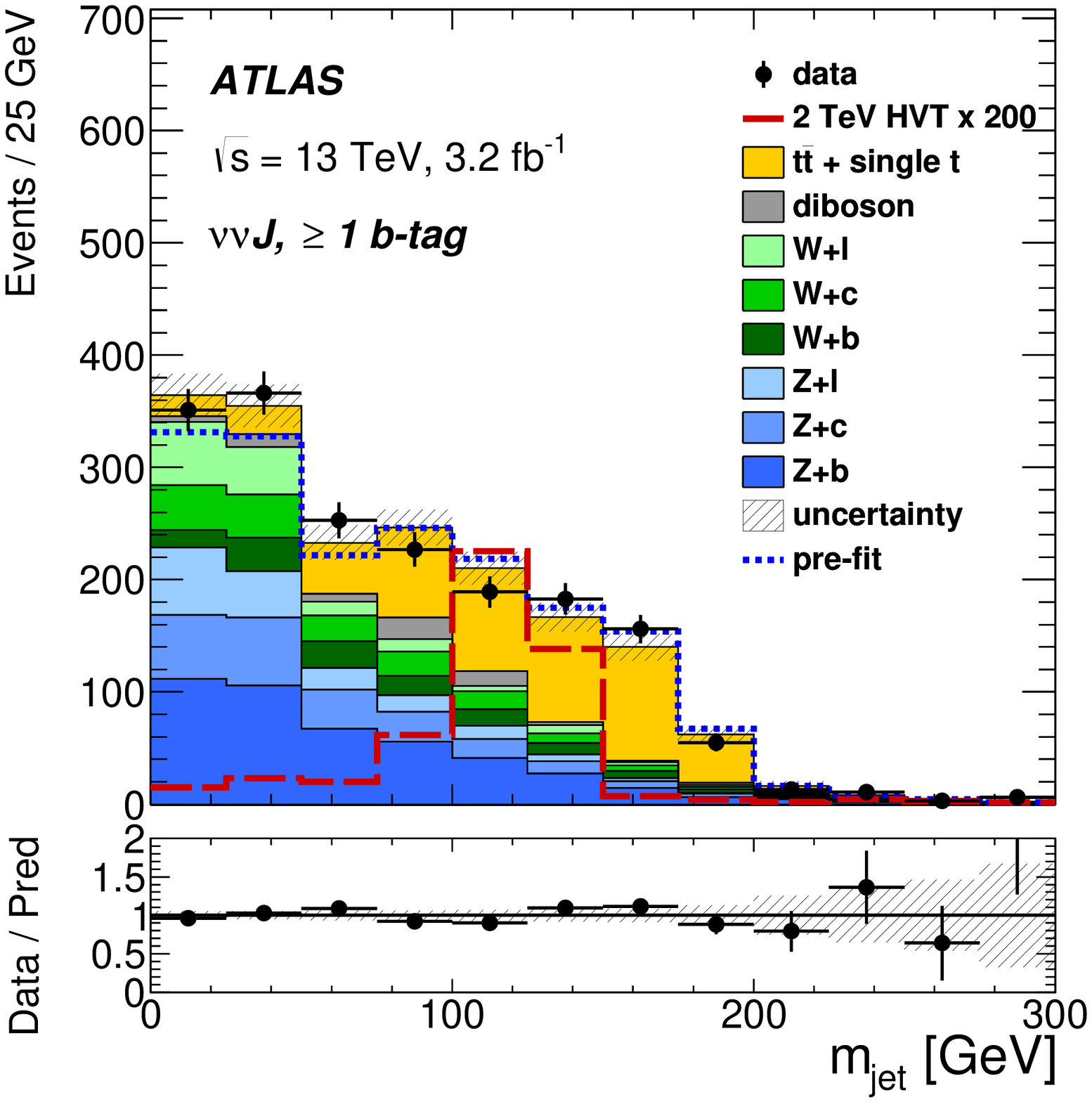}
}
\subfigure[][1-lepton, $\geq 1$ $b$-tag]{
  \includegraphics[width=0.45\textwidth]{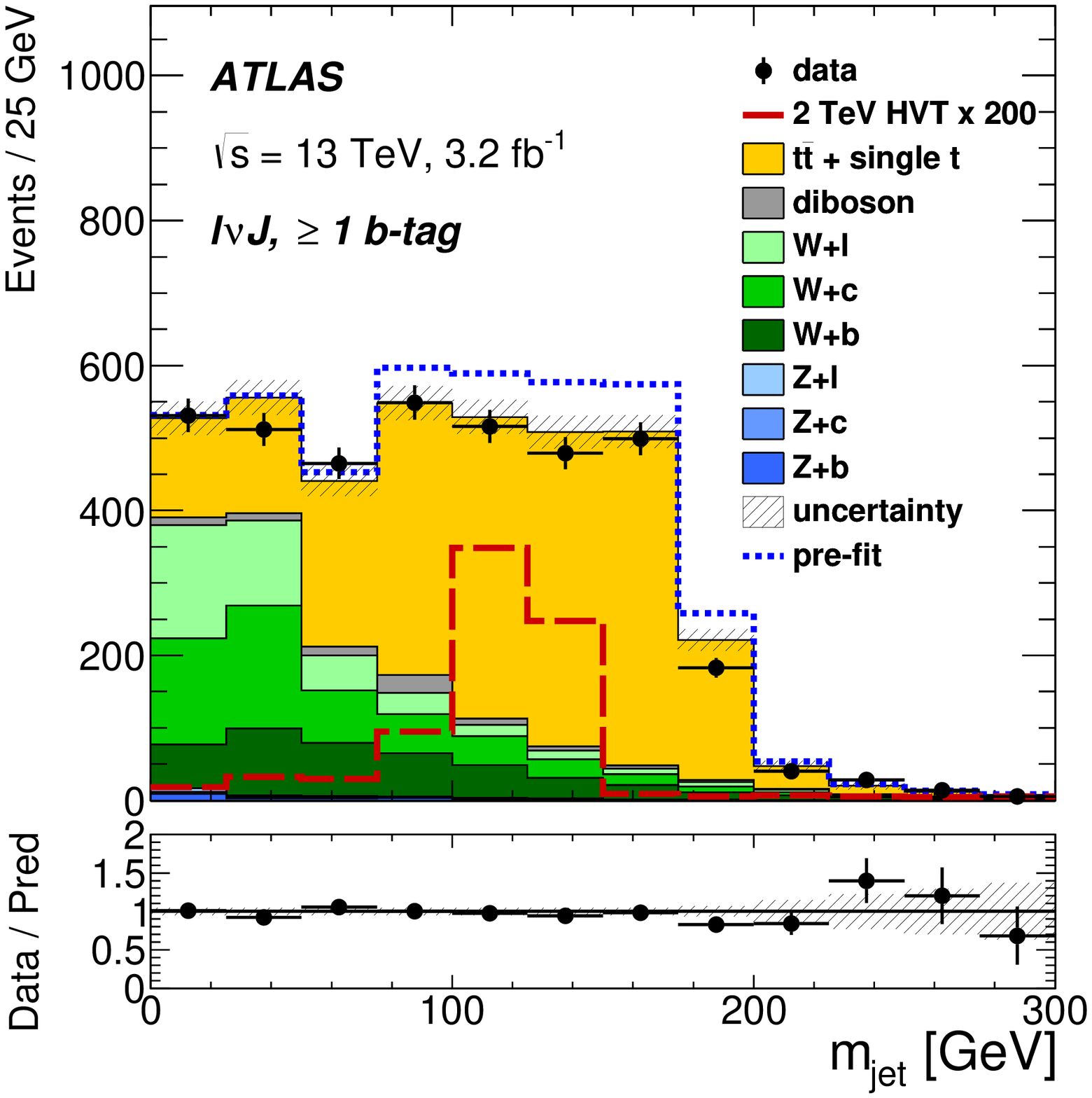}
}
\subfigure[][2-lepton, $\geq 1$ $b$-tag]{
    \includegraphics[width=0.45\textwidth]{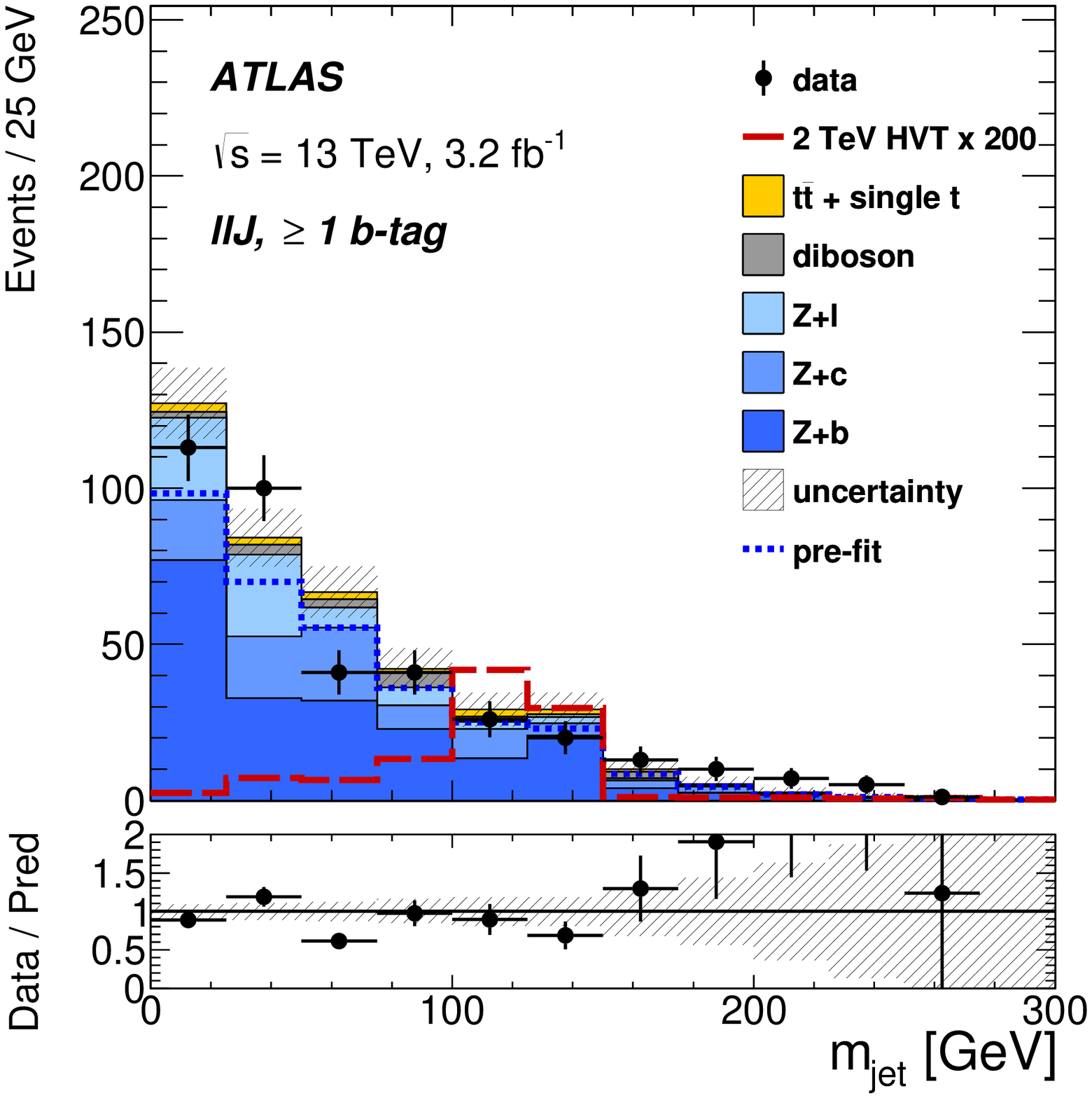}
}
  \caption{ Distributions of the mass of the leading large-$R$ jet,
      $m_{\rm jet}$, for
      the (a) 0-lepton, (b) 1-lepton, and (c) 2-lepton channels.
      The background prediction is shown after the maximum-likelihood
      fits to the data described in Section~\ref{sec:results}; the total
      background prediction before the fit is shown by the dotted
      blue line.
      The signal for the benchmark HVT \textit{Model A} with $m_{V'} =
      2$~TeV is shown as a dotted red line and normalized to 200 times
      the theoretical cross-section.
	}
	  \label{fig:allbb_final_mass}
	  	\end{figure}

The four-momentum of the large-$R$ jet is corrected by adding the
four-momentum of the muon closest in $\Delta R$ to the jet axis
provided it is within the jet radius.
The distribution of the mass of the leading large-$R$ jet ($m_{\rm
jet}$) in events passing the selection described so far is shown in
Figure~\ref{fig:allbb_final_mass}.
The mass of the leading large-$R$ jet (${\rm jet}$) is required to be consistent
with the Higgs boson mass of 125.5~GeV.
A 90\% efficient mass requirement, corresponding to a window of 75 GeV
$<m_{{\rm jet}}<$ 145 GeV, is applied.
This is particularly effective for discriminating the signal from
$t\bar{t}$ and $V+b \bar{b}$ backgrounds. 

The events passing this selection, and categorized into 0-, 1-, and
2-lepton channels by 1- and 2-$b$-tags (six categories in total),
define the signal regions of this analysis.
The efficiencies of selecting events in the 2-$b$-tag (1-$b$-tag)
signal region for an HVT resonance of mass of 1.5 TeV are 24\%
(34\%),  16\% (25\%) and 15\% (22\%) for the $Z^\prime \rightarrow Z H \rightarrow \nu \bar{\nu}
b\bar{b}$ , $W^\prime \rightarrow W H
\rightarrow \ell \nu b\bar{b}$  and $Z^\prime \rightarrow  Z H\rightarrow
\ell^+ \ell^- b\bar{b}$ processes respectively.

%-------------------------------------------------------------------------------
\section{Background estimation}
\label{sec:background}
%-------------------------------------------------------------------------------
\label{backgrounds}

The background contamination in the signal regions is different for
each of the three channels.
In the 0-lepton analysis the dominant background is $Z$+jets
production with significant contributions from $W$+jets and $t\bar{t}$
production.
In the 1-lepton channel the dominant backgrounds are $W$+jets and $t\bar{t}$ production.
In the 2-lepton channel, where two same-flavour leptons with an
invariant mass near the $Z$ mass are selected, $Z$+jets production
is by far the dominant background.
All three channels also have  small contributions from
single-top-quark, diboson and SM Higgs production.
The multi-jet background, which enters the signal regions through semileptonic
hadron decays and through misidentified or mismeasured jets, is
found to be negligibly small in all three channels.

The background modelling is studied using control regions with low
signal contamination, chosen to not overlap with the signal regions.
These control regions are used both to evaluate the background
predictions outside the signal-rich regions and to establish the
normalization and $m_{\rm VH}$ shape of the dominant backgrounds
through their inclusion as nuisance parameters in the likelihood fit
described in Section~\ref{sec:results}.

Sideband regions of the $m_{\rm jet}$ distribution, defined as
$m_{{\rm jet}}<$ 75 GeV (low-$m_{\rm jet}$) or $m_{{\rm jet}}>$ 145
GeV (high-$m_{\rm jet}$) are used as control regions for the
$W/Z$+jets backgrounds.
Furthermore, the events are divided into categories corresponding to
the number of $b$-tagged track jets matched to the large-$R$ jet to
test the different flavour compositions.
The 1- and 2-$b$-tag low-$m_{\rm jet}$ control regions mainly test the
$W/Z+c$ and  $W/Z+b$  contributions, respectively.

Control regions for the $t \bar{t}$ background prediction are also defined.
For the 0- and 1-lepton channels, the $t \bar{t}$ control regions are
defined by requiring at least one additional $b$-tagged track jet that
is not matched to the large-$R$ jet; no Higgs boson candidate mass window
requirement is imposed in the 0- and 1-lepton \ttbar\ control regions.
The $t \bar{t}$ control region for the 2-lepton channel is defined by
requiring exactly one electron, exactly one muon and at least one
$b$-tagged track jet matched to the leading large-$R$ jet; there is
no requirement on additional $b$-tagged track jets in the 2-lepton
channel.

%-------------------------------------------------------------------------------
\section{Systematic uncertainties}
\label{sec:syst}
%-------------------------------------------------------------------------------
\label{systematics}

The most important experimental systematic uncertainties are
associated with the measurement of the scale and resolution of the
large-$R$ jet energy and mass, as well as with the determination of
the track jet $b$-tagging efficiency and mistag rate.
The uncertainties in the scale and resolution of large-$R$ jet energy
and mass are evaluated by comparing the ratio of calorimeter-based to
track-based measurements in multi-jet data and
simulation~\cite{ATL-PHYS-PUB-2015-033}. 
The uncertainty in the track-jet $b$-tagging  efficiency arises
mainly from uncertainty in the measurement of the $b$-tagging
efficiency in $t\bar{t}$ events, while the mistag rate and uncertainty
are determined using dijet events~\cite{Aad:2015ydr}.
These uncertainties have an impact on the normalization and
differential distribution of events, and have typical sizes of 2--20\%
for the large-$R$ jet energy/mass scales and 5--15\% for the
$b$-tagging efficiency.

Other experimental systematic uncertainties with a smaller impact are
those associated with the lepton energy and momentum scales, lepton
identification efficiency, the efficiency of the triggers, 
the small-$R$ jet energy scale and the \met measurement.

Uncertainties are taken into account for possible differences
between data and the simulation model that is used for each process.
In addition to the 5\% uncertainty in the integrated luminosity, the
following normalization uncertainties are assigned to particular
processes: 30\% for $t\bar{t}$ and
single top quarks ~\cite{Aad:2015hna}, 11\% for
dibosons~\cite{Campbell:2010ff}, 10\% for $W/Z+$light
jets~\cite{Aad:2014qxa}, and 30\% for $W/Z+c$ and $W/Z+b$.
Uncertainties in the modelling of the $m_{VH}$ and $m_{{\rm T}, VH}$
distributions are assigned to the $Z$+jets and $W$+jets backgrounds.
These uncertainties are estimated by comparing predictions from {\sc
Sherpa 2.1.1} and MadGraph5\_aMC@NLO-2.2.2 at leading order with
showering by {\sc Pythia} 8.186 using the A14 tune.
An uncertainty in the shape of the  $m_{VH}$ or $m_{{\rm T}, VH}$
distribution for the $t\bar{t}$  background is derived by comparing a
{\sc Powheg} sample with the distribution obtained using
MadGraph5\_aMC@NLO 2.2.2.
Additional systematic uncertainties are evaluated by comparing the
nominal sample showered with {\sc Pythia} 6.428 using the P2012 tune to
one showered with {\sc Herwig++} 2.7.1~\cite{Herwig} and using the
UEEE5 underlying-event tune.
Samples of $t \bar{t}$ events with the factorization and renormalization
scale doubled or halved are compared to the nominal, and differences
observed are taken as an additional uncertainty.

The dominant uncertainties in the signal acceptance arise from the choice
of PDF and from uncertainty in the amount of initial- and final-state
radiation present in simulated signal events.
The PDF uncertainties are estimated by taking the acceptance
difference between the NNPDF2.3LO and MSTW2008LO PDF
and adding it in quadrature with the differences in acceptance found
between the NNPDF2.3LO error sets.
Typical values for the signal acceptance uncertainties are 2--3\% per
source of uncertainty.

All uncertainties are evaluated in an identical way for all signal
and background sources and are thus treated as fully correlated
across sources.
For all simulated samples, the statistical uncertainty arising from
the limited number of simulated events is taken into account.

%-------------------------------------------------------------------------------
\section{Results}
\label{sec:result}
\label{sec:results}

\begin{figure}[tb]
\centering

\subfigure[][0-lepton, 1-$b$-tag]{
  \includegraphics[width=0.45\textwidth]{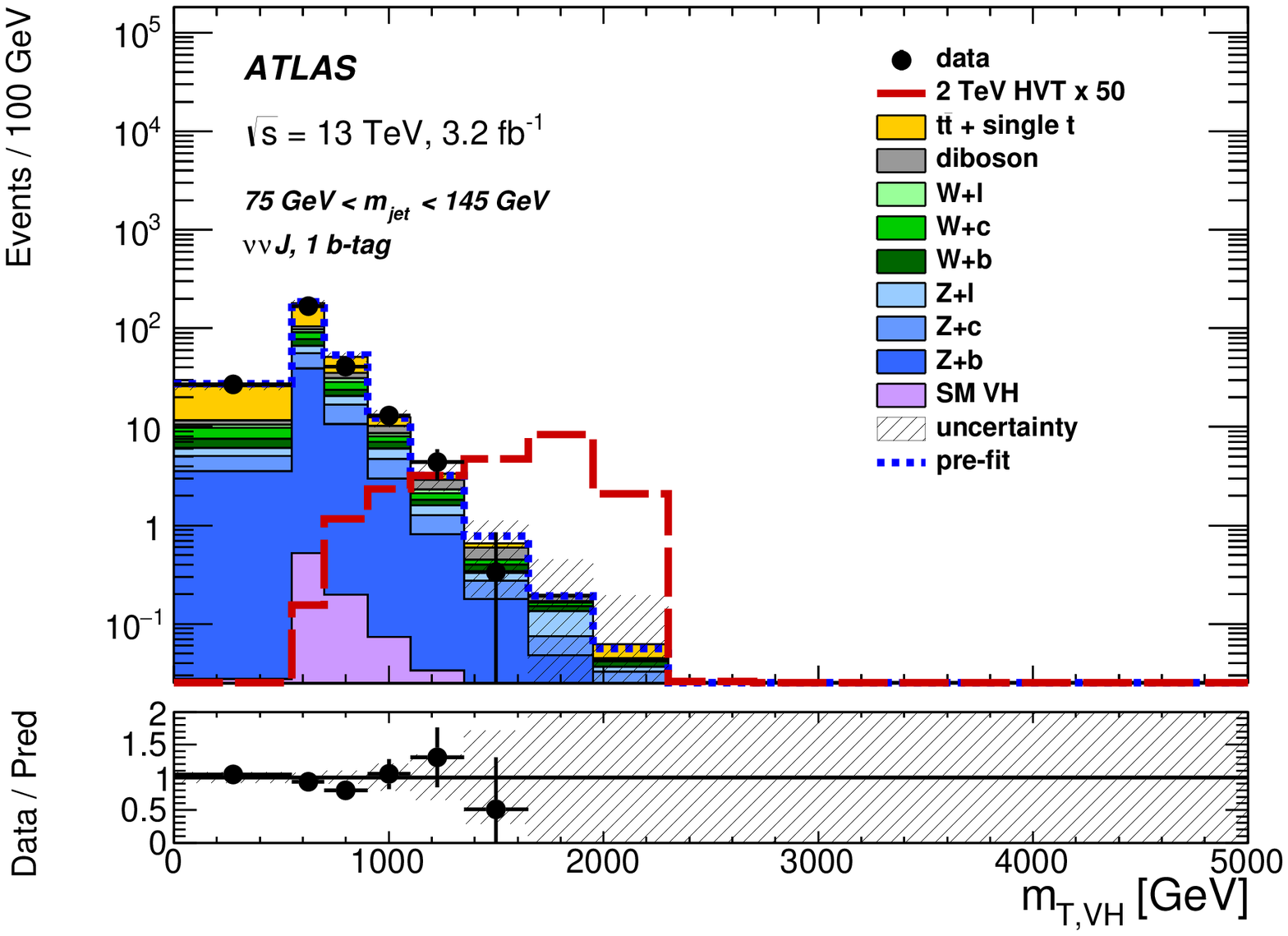}
}
\subfigure[][0-lepton, 2-$b$-tag]{
  \includegraphics[width=0.45\textwidth]{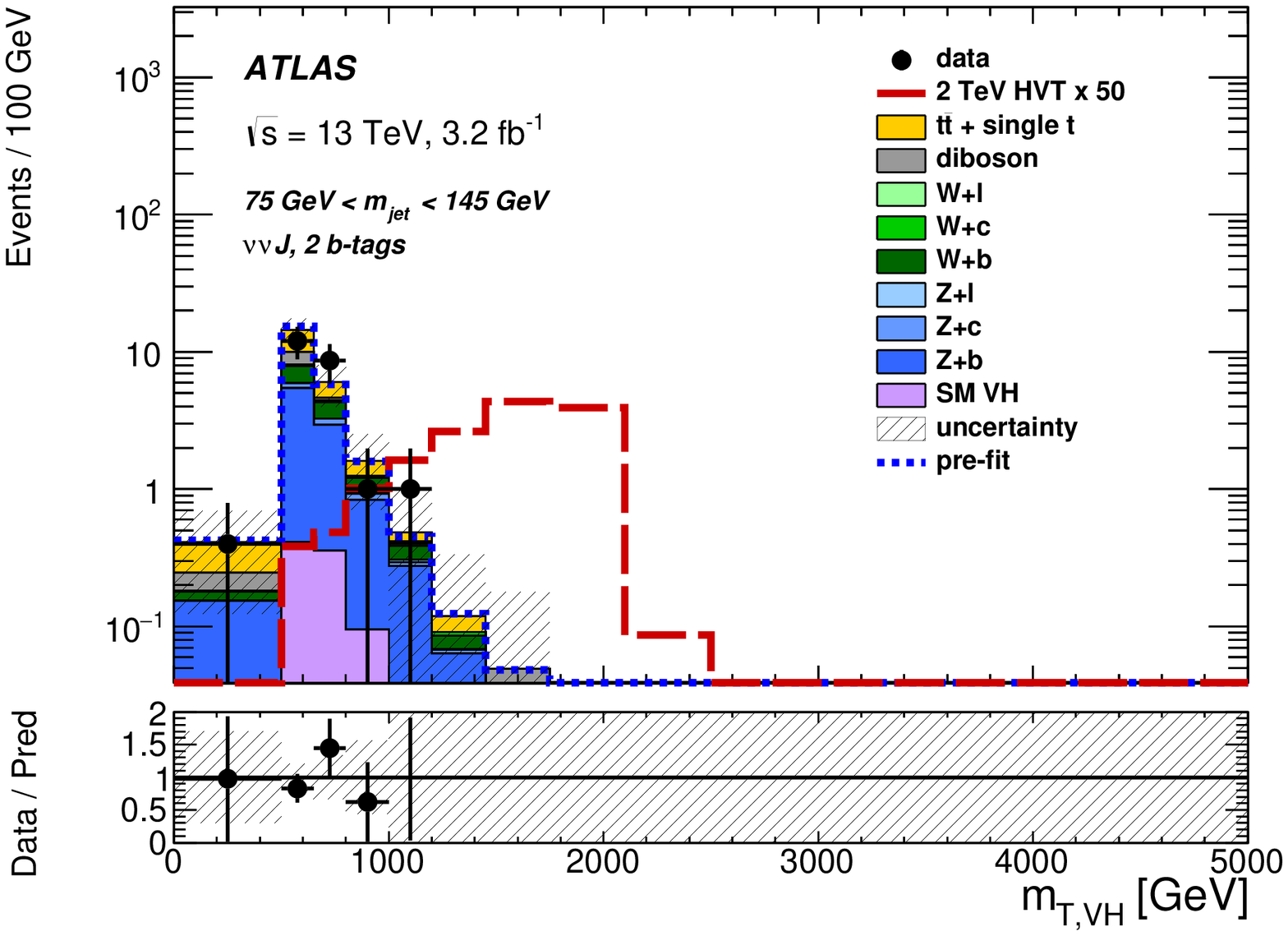}
}
\\
\subfigure[][1-lepton, 1-$b$-tag]{
  \includegraphics[width=0.45\textwidth]{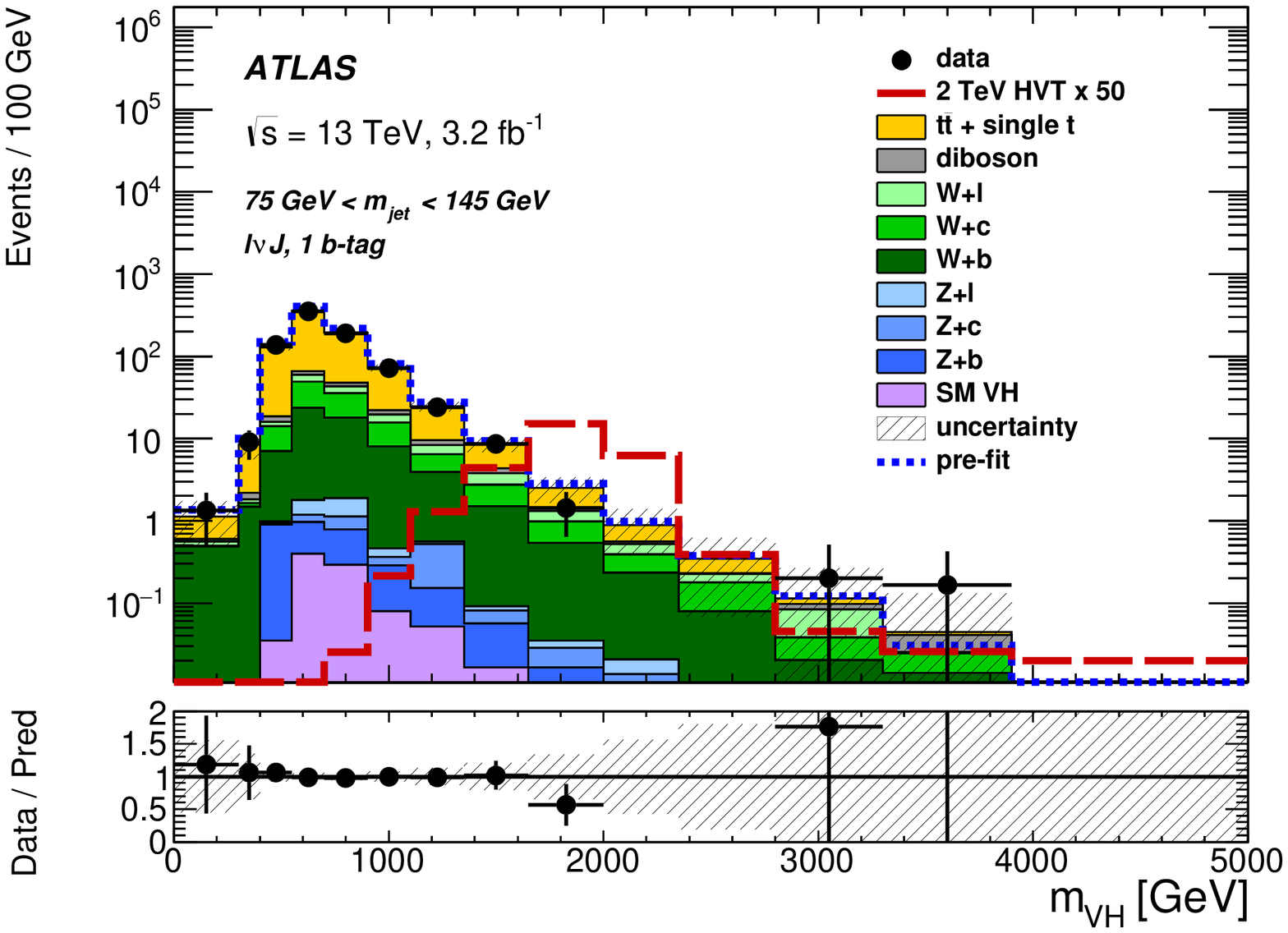}
}
\subfigure[][1-lepton, 2-$b$-tag]{
  \includegraphics[width=0.45\textwidth]{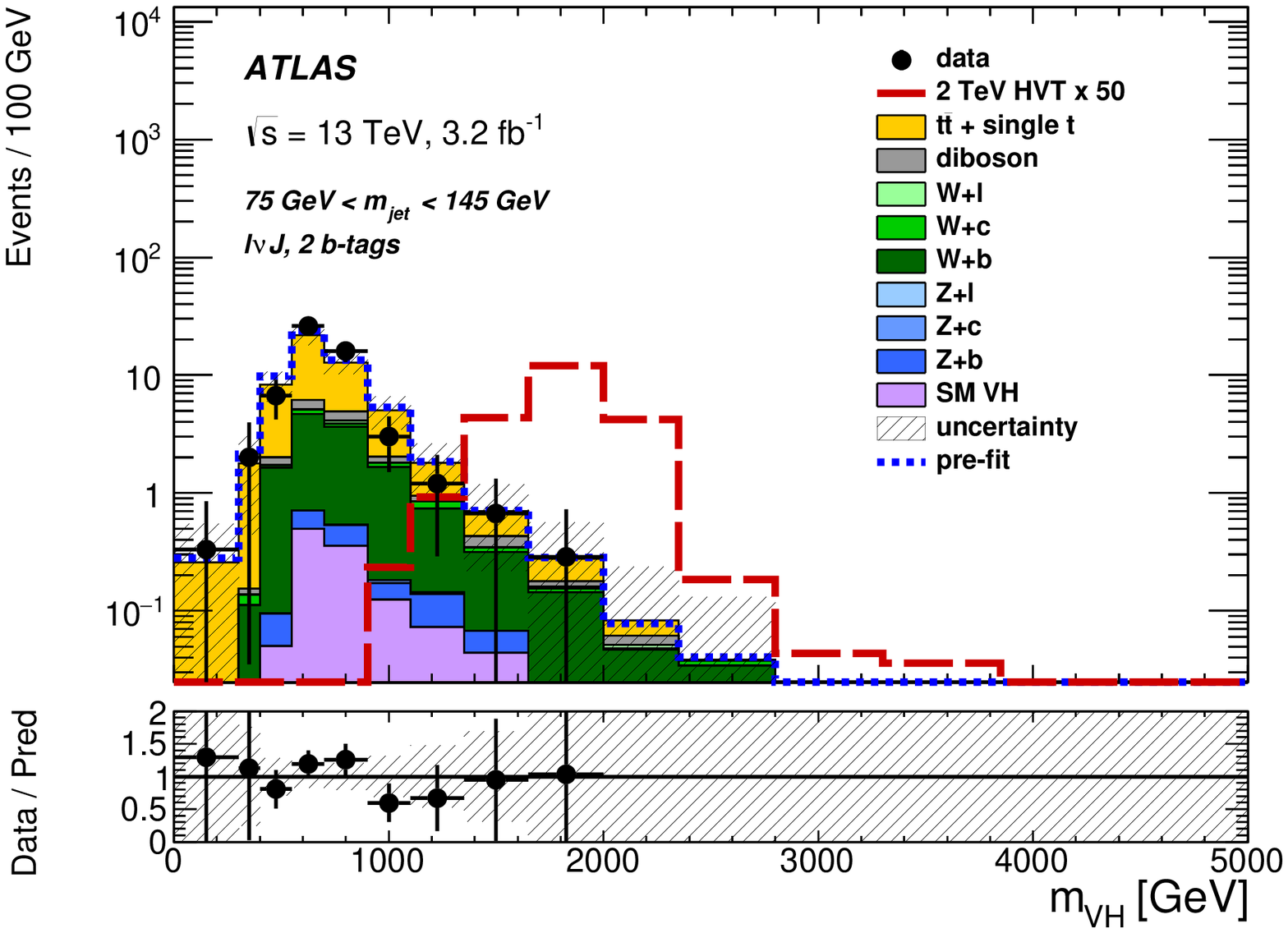}
}
\\
\subfigure[][2-lepton, 1-$b$-tag]{
  \includegraphics[width=0.45\textwidth]{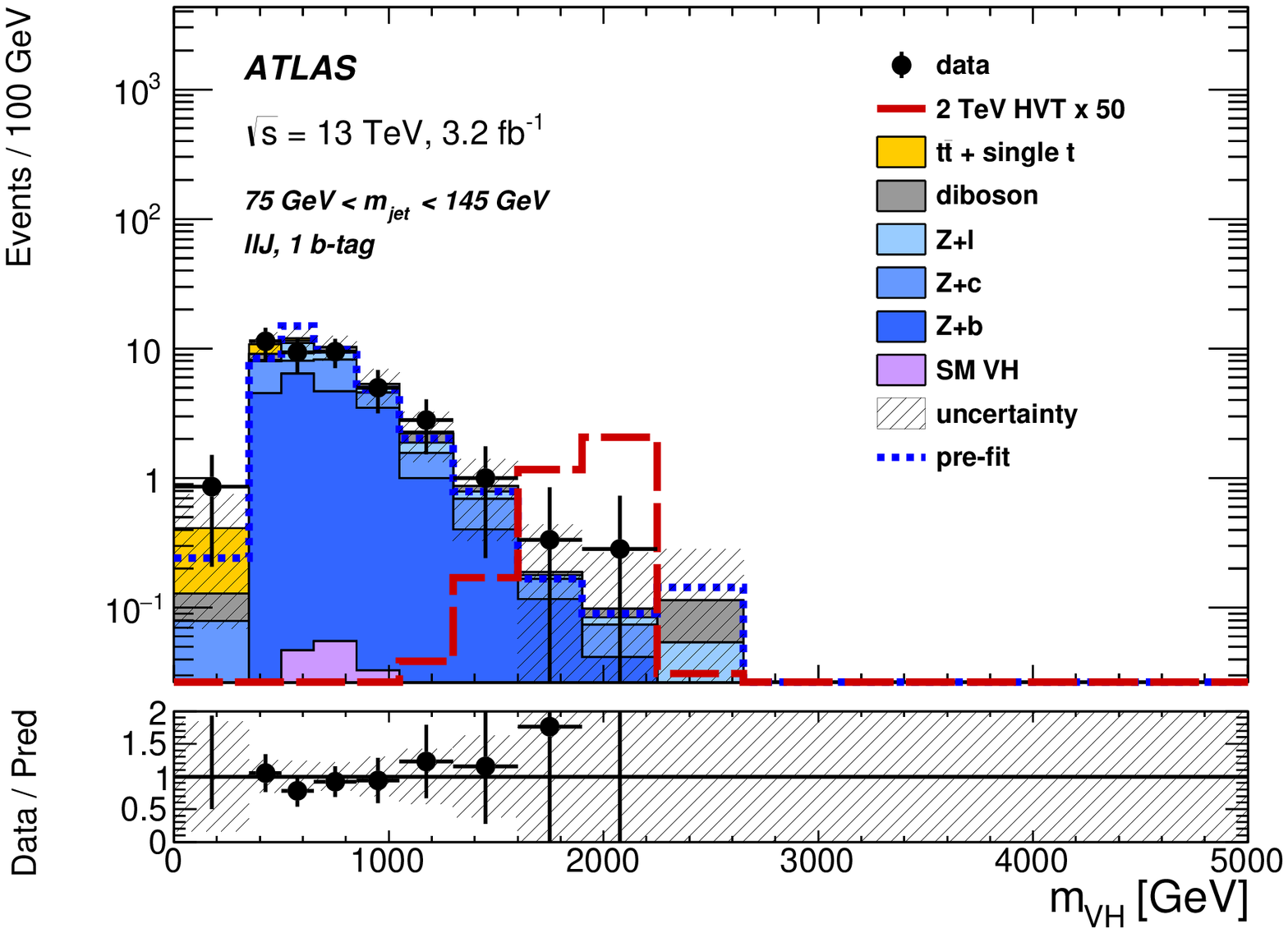}
}
\subfigure[][2-lepton, 2-$b$-tag]{
  \includegraphics[width=0.45\textwidth]{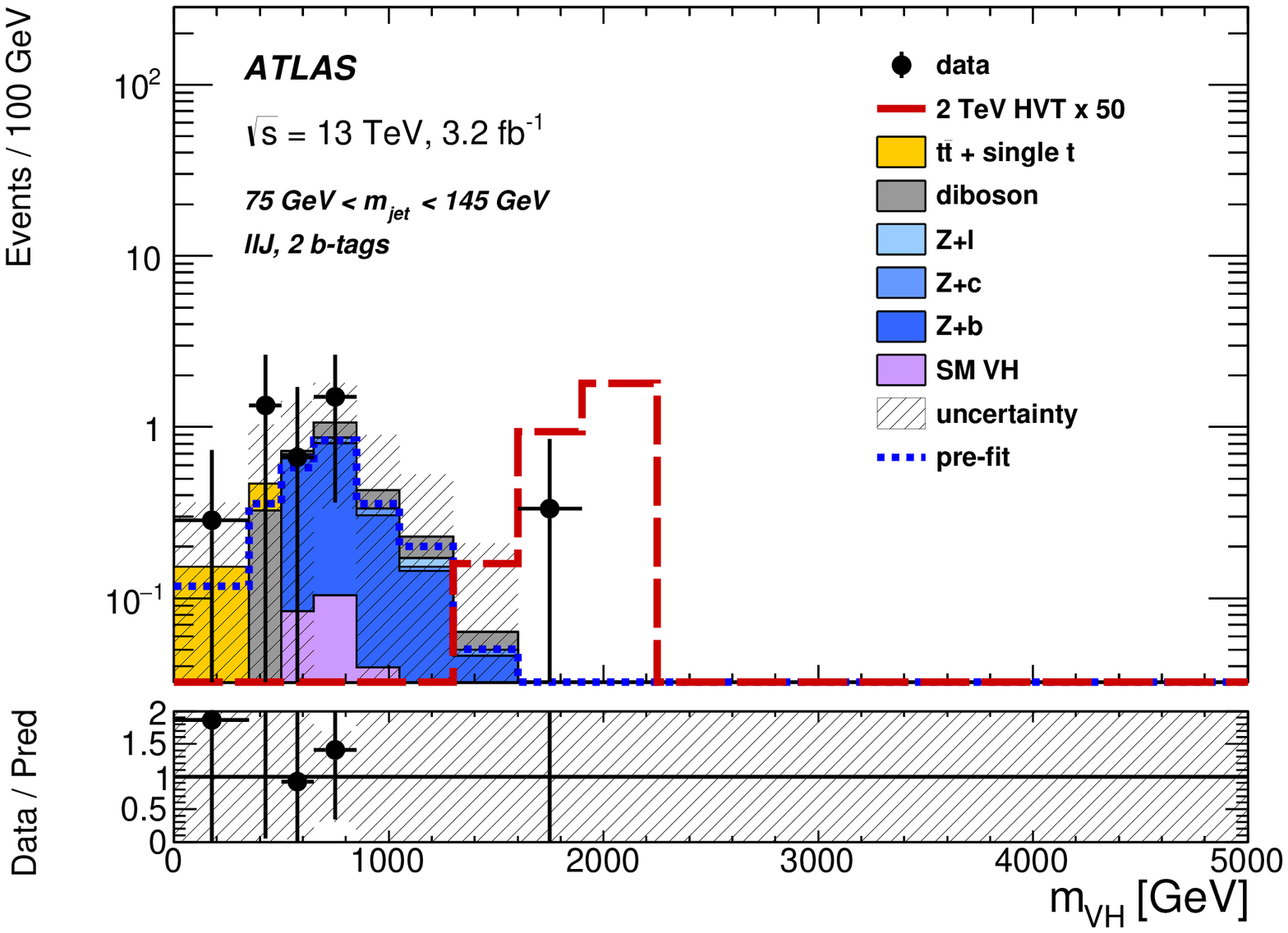}
}

\caption{ Distributions of reconstructed $VH$ transverse mass,
    $m_{{\rm T}, VH}$, and invariant mass, $m_{VH}$, for the
    0-lepton (top), 1-lepton (middle), and 2-lepton (bottom)
    channels.
    The left (right) column corresponds to the 1-$b$-tag (2-$b$-tag)
    signal regions.
    The background prediction is shown after the maximum-likelihood
    fits to the data; the total background prediction before the
    fit is shown by the dotted blue line.
    The signal for the benchmark HVT \textit{Model A} with $m_{V'} =
    2$~TeV is shown as a dotted red line and normalized to 50 times
    the theoretical cross-section.
    } \label{fig:xxbb_final}
\end{figure}

\begin{table}
\caption{The predicted and observed number events for the three
    final states considered in this analysis.
    The predicted number of events is shown after a maximum-likelihood
    fit to the data, performed simultaneously across the three lepton
    channels.
    The quoted uncertainties are the combined systematic and
    statistical uncertainties after the fit.
    Uncertainties in the normalization of individual backgrounds may
    be larger than the uncertainty on the total background due to
    correlations.}
\centering
\begin{footnotesize}
\sisetup{
    table-number-alignment = center,
    table-figures-integer = 2,
    group-digits = false
}
\begin{tabular}{l
                S[separate-uncertainty,table-figures-uncertainty=1]
                S[separate-uncertainty,table-figures-uncertainty=1]
                S[separate-uncertainty,table-figures-uncertainty=1]
}
\toprule

\multicolumn{4}{c}{Two $b$-tags} \\
& $\nu\nu b \bar{b}$
& $\ell\nu b \bar{b}$
& $\ell\ell b \bar{b}$ \\
\midrule
$t\bar{t}$      & 9.6(14)  & 50(7)    & 0.54(36) \\
Single top      & 2.0(6)   & 11.4(30) & 0.20(10) \\
$W+b$           & 5.2(13)  & 18(5)    &          \\
$W+c$           & 0.64(18) & 2.0(7)   &          \\
$W+q$           & 0.06(3)  & 2.0(8)   &          \\
Diboson         & 4.2(18)  & 4.6(8)   & 1.28(27) \\
SM $VH$         & 1.43(57) & 0.03(1)  & 0.45(19) \\
$Z+b$           & 12.3(24) & 1.0(4)   & 3.4(8)   \\
$Z+c$           & 1.46(43) & 0.05(2)  & 0.31(10) \\
$Z+q$           & 0.13(5)  &          & 0.04(4)  \\
\midrule                                                                                                     
Backgrounds     & 36.9(35) & 90(6)    & 6.2(10)  \\
Data            & 37       & 96       & 8        \\
\midrule
\midrule

\multicolumn{4}{c}{One $b$-tag} \\
& $\nu\nu b \bar{b}$
& $\ell\nu b \bar{b}$
& $\ell\ell b \bar{b}$ \\
\midrule
$t\bar{t}$        & 216(17) & 969(50)    & 3.8(8)   \\
Single top        & 26(7)   & 112(30)    & 0.58(19) \\
$W+b$             & 33(8)   & 100(24)   &          \\
$W+c$             & 41(10)  & 109(31)   &          \\
$W+q$             & 20(5)   & 1.5(6)    &          \\
Diboson           & 28(5)   & 32(5)     & 6.4(10)  \\
SM $VH$           & 1.6(6)  & 0.04(1)   & 0.30(12) \\
$Z+b$             & 99(17)  & 3.8(10)   & 36(6)    \\
$Z+c$             & 51(13)  & 2.7(16)   & 19(5)    \\
$Z+q$             & 32(8)   & 3.0(10)   & 9(4)     \\
\midrule                                                                                                                            
Backgrounds       & 548(16) & 1385 (30) & 75(7)    \\
Data              & 520     & 1364      & 75       \\
\bottomrule

\end{tabular}       
\end{footnotesize}
\label{tab:yield}
\end{table}

To determine how well the observed data agrees with the predicted
backgrounds and to test for an HVT signal, a maximum-likelihood fit is
performed over the binned $m_{VH}$ or $m_{{\rm T}, VH}$ mass
distributions, including all control regions described in
Section~\ref{backgrounds}.
The maximum-likelihood fit parameters are the systematic uncertainties
in each background and signal contribution, which can vary the
normalizations and differential distributions.  The systematic
uncertainties are given log-normal priors in the likelihood, with
scale parameters described in Section~\ref{systematics}.
High- and low-$m_{\rm jet}$ sideband control regions are merged if fewer
than 100 background events are expected with the full dataset; this
is the case for the 0-lepton 2-$b$-tag\ sidebands, the 1-lepton
2-$b$-tag sidebands, and the 2-lepton 1- and 2-$b$-tag sidebands.
The HVT signal is included as a binned template with an unconstrained
normalization.

Table~\ref{tab:yield} provides the predicted and observed number of
events in each signal region, and the reconstructed mass distributions
for events passing the selections are shown in
Figure~\ref{fig:xxbb_final}.
The predicted background is shown after the binned maximum-likelihood
fit to the data, performed simultaneously across lepton channels.

No significant excess of events is observed in the data compared to
the prediction from SM background sources.
Exclusion limits at the 95\% confidence level are set on the
production cross-section times the branching fraction for the HVT
models. The limits for the charged resonance, $W^{\prime}$,  are
obtained by performing the likelihood fit over the 0- and 1-lepton
channels, while the 0- and 2-lepton channels are used for the neutral
resonance, $Z^{\prime}$.
In the case of the $W^{\prime}$ search, the $\tau$-lepton veto is not
imposed and the search considers only the $W^{\prime}\rightarrow
WH$ signal, while for the $Z^{\prime}$ search the $\tau$ veto is
imposed and only $Z^{\prime}\rightarrow ZH$ signal is considered.

The results for combined HVT production are evaluated without the
$\tau$ veto imposed, including both the $W^{\prime}\rightarrow WH$
and $Z^{\prime}\rightarrow ZH$ signals simultaneously.
The combined HVT $V^\prime$ search is performed with
maximum-likelihood fits that are independent from those of the
$W^\prime$ and $Z^\prime$ searches, so there is no double-counting of
0-lepton events that are included in the individual fits.

\begin{figure*}[ht!]
    \centering

    \subfigure[$Z'  \rightarrow ZH$, $H \rightarrow b\bar{b}/c\bar{c}$]{
        \includegraphics[width=0.45\textwidth]{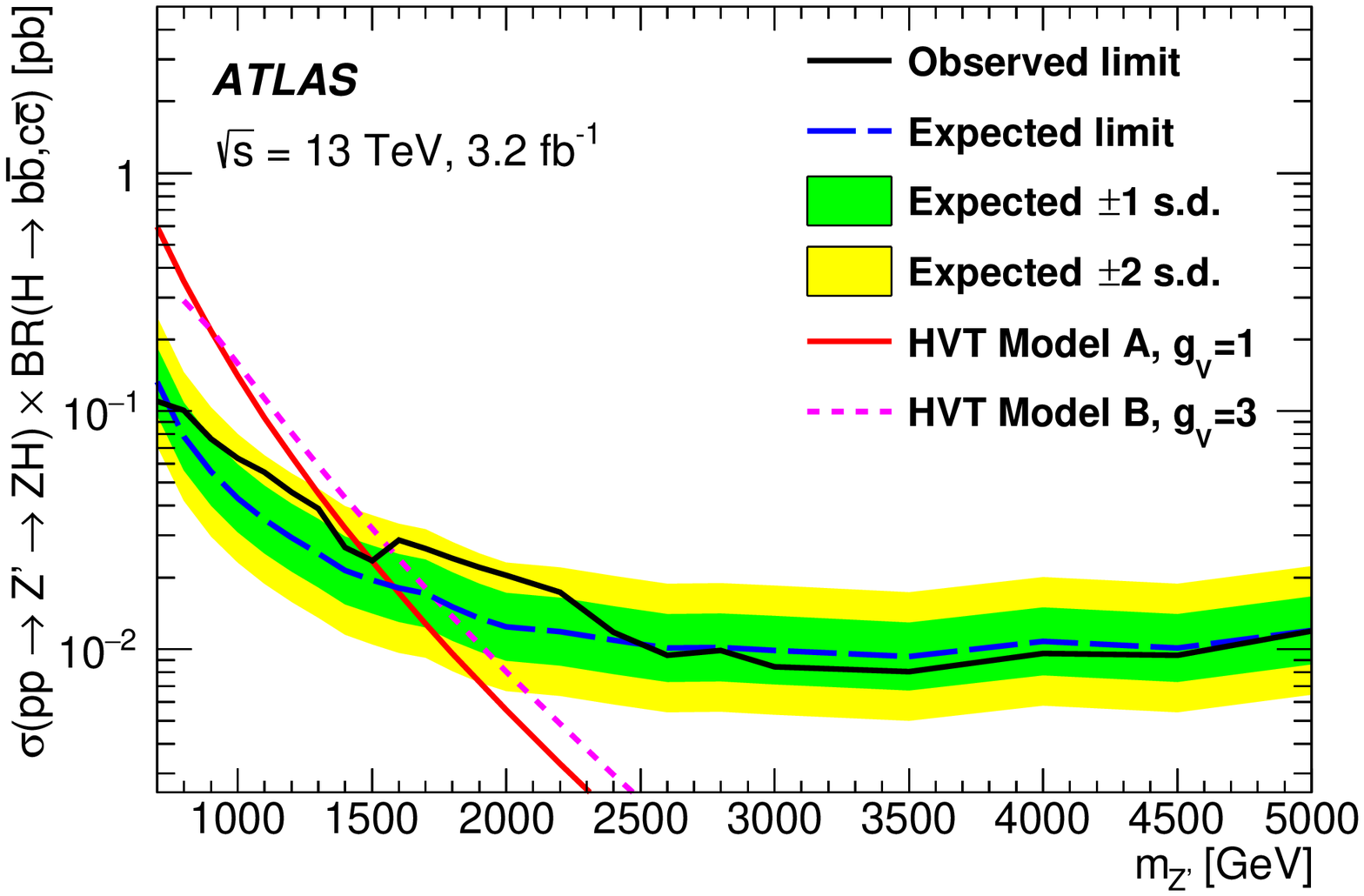}
    }
    \subfigure[$W'^{\pm}  \rightarrow WH$, $H \rightarrow b\bar{b}/c\bar{c}$]{
        \includegraphics[width=0.45\textwidth]{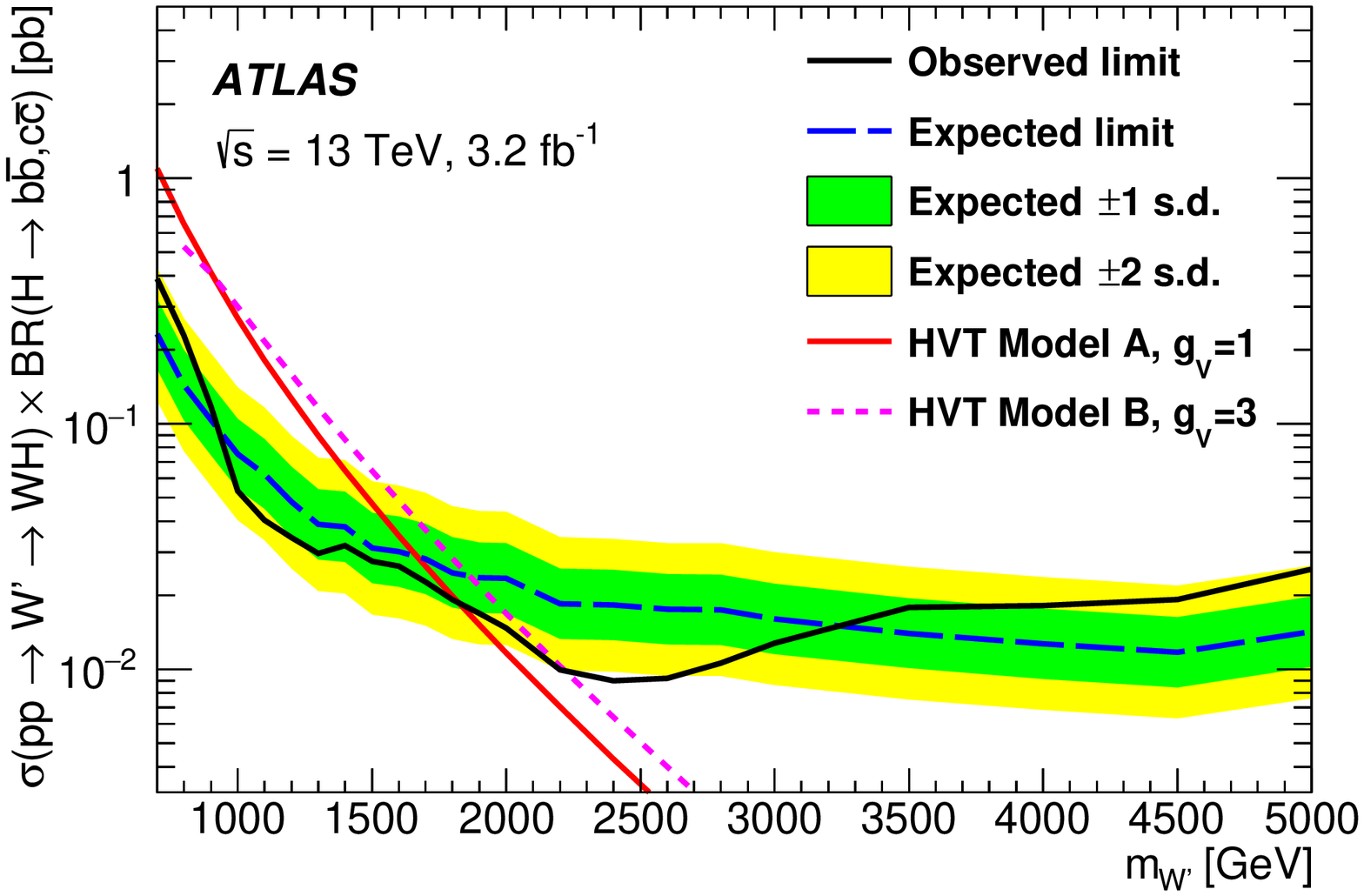}
    }\\

    \subfigure[$V'^{\pm}  \rightarrow WH/ZH$] {
        \includegraphics[width=0.45\textwidth]{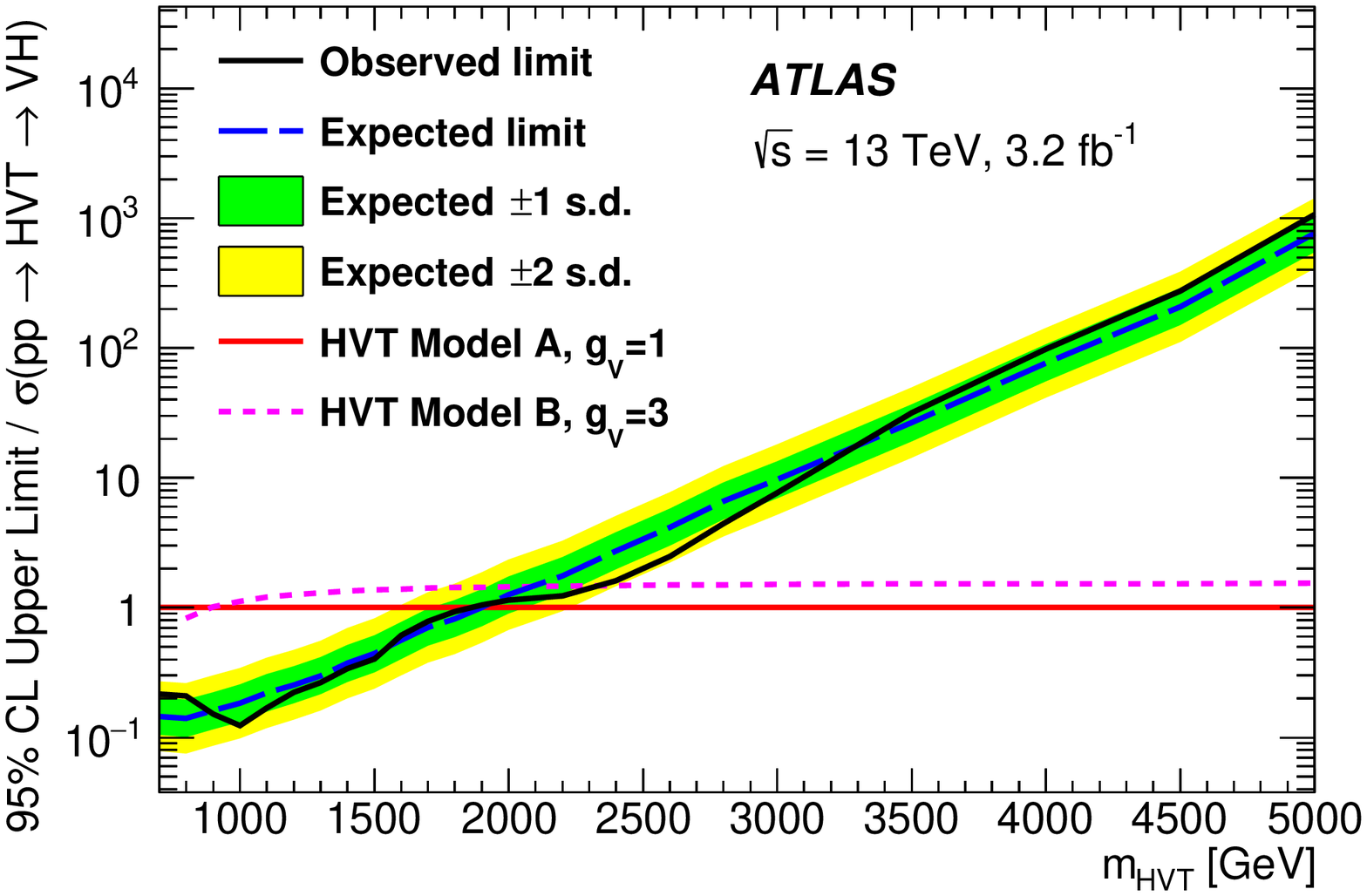}
    }
    \caption{Upper limits at the 95\% CL for (a) the production
        cross-section of $Z^{\prime}$ times its branching fraction to
        $ZH$ and the branching ratio BR$(H \rightarrow
        b\bar{b}/c\bar{c})$ and (b) the production cross-section of
        $W^{\prime}$ times its branching fraction to $WH$ and the
        branching ratio BR$(H \rightarrow b\bar{b}/c\bar{c})$.
        Upper limits at the 95\% CL for (c) the scaling factor of the
        production cross-section for $V'$ times its branching fraction
        to $WH/ZH$ in \textit{Model A}.
        The production cross-sections predicted by \textit{Model A}
        and \textit{Model B} are shown for comparison.
        } \label{fig:limit}
\end{figure*}

The exclusion limits are calculated with a modified frequentist
method \cite{CLs_2002}, also known as \textit{CL$_{s}$},
and the profile-likelihood-ratio test statistic \cite{Cowan2011} in
the asymptotic approximation, using the binned $m_{VH}$ or $m_{{\rm
T}, VH}$ mass distributions for 0-, 1- and 2-lepton final states.
Systematic uncertainties and their correlations are taken into account
as nuisance parameters. None of the systematic uncertainties
considered are significantly constrained or pulled in the likelihood
fit.
Figures~\ref{fig:limit}(a) and ~\ref{fig:limit}(b) show the 95\% CL
upper limits on the production cross-section multiplied by the
branching fraction into $WH$ and $ZH$ and the branching fraction BR$(H
\rightarrow b\bar{b}/c\bar{c})$ as a function of the resonance mass,
separately for the charged $W^{\prime}$ and the neutral $Z^{\prime}$
bosons, respectively.
The theoretical predictions for the HVT benchmark \textit{Model A}
with coupling constant $g_{V}=1$ allow exclusion of $m_{Z^{\prime}} <
1490$~GeV and $m_{W^{\prime}} < 1750$~GeV.
For \textit{Model B} with coupling constant $g_{V}=3$ the
corresponding excluded masses are
$m_{Z^{\prime}} < 1580$~GeV  and $m_{W^{\prime}} < 2220$~GeV.
In both theoretical predictions, the branching fraction
BR$(H \rightarrow b\bar{b}/c\bar{c})$ is fixed to the Standard Model
prediction of 60.6\%~\cite{Heinemeyer:2013tqa}.

To study the scenario in which the masses of charged and neutral
resonances are degenerate, a combined likelihood fit over all the
signal regions and control regions is also performed.
The  95\% CL upper limits on the production cross-section of a $V'$
decaying to $WH/ZH$, relative to the HVT model predictions, are shown
in Figure~\ref{fig:limit}(c).
For \textit{Model A} (\textit{Model B}) with coupling constant $g_V =
1$ ($g_V = 3$), $m_{V'^\pm} < 1730$~GeV (2310~GeV) is excluded.

\begin{figure}
    \centering	
    \includegraphics[width=0.48\textwidth]{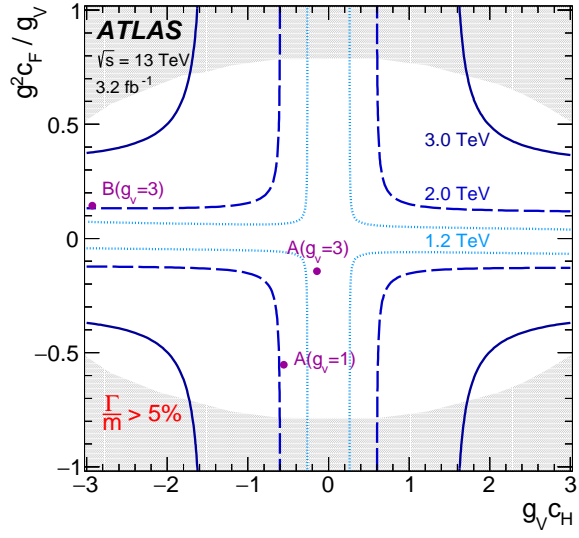}	    
    \caption{Observed 95\% CL exclusion contours in the HVT parameter
    space \{$g_Vc_H$, ($g^2$/$g_V$)$c_F$\} for resonances of mass
    1.2~\TeV, 2.0~\TeV\ and 3.0~\TeV. The areas outside the curves are
    excluded.  Also shown are the benchmark model parameters A($g_V$=1),
    A($g_V$=3) and B($g_V$=3).
    The shaded region corresponds to the parameter values for which
    the resonance total width $\Gamma$ is greater than 5\% of its
    mass, in which case it is not negligible compared to the
    experimental resolution.}
        \label{fig:HVT_contours}
\end{figure}

The exclusion contours in the HVT parameter space \{$g_Vc_H$,
($g^2$/$g_V$)$c_F$\} for resonances of mass 1.2~\TeV, 2.0~\TeV\ and
3.0~\TeV\ are shown in Figure~\ref{fig:HVT_contours} where all three
channels are combined, taking into account the branching fractions to
$WH$ and $ZH$ from the HVT model parameterization. 
Here the parameter $c_F$ is assumed to be the same for quarks and
leptons, including third-generation fermions, and other parameters
involving more than one heavy vector boson, $g_Vc_{VVV}$,  $g_V^2
c_{VVHH}$ and $c_{VVW}$, have negligible contributions to the overall
cross-sections for the processes of interest.

%-------------------------------------------------------------------------------

\FloatBarrier

%-------------------------------------------------------------------------------
\section{Conclusion}
\label{sec:conclusion}
%-------------------------------------------------------------------------------

A search for a new, heavy resonance decaying to $WH/ZH$ is presented.
The search is performed using  $3.2\pm0.2$ fb$^{-1}$ of $pp$ collision
data at  a 13 TeV centre-of-mass energy collected by the ATLAS
detector at the Large Hadron Collider.
No significant deviations from the SM background predictions are
observed in the three final states considered: $\ell^+ \ell^- b \bar{b}$,
$\ell \nu b \bar{b}$, $\nu \bar{\nu} b \bar{b}$.
Upper limits are set at
the 95\% confidence level on the production cross-sections of $V^\prime$
in heavy vector triplet models with resonance masses above 700~GeV.
HVT benchmark \textit{Model A} with coupling constant $g_{V}=1$ is
excluded for $m_{Z^{\prime}}<1490$~GeV, $m_{W^{\prime}} < 1750$~GeV,
and $m_{V^\prime}<1730$~GeV; for \textit{Model B} with coupling
constant $g_{V}=3$, $m_{Z^{\prime}}<1580$~GeV, $m_{W^{\prime}} <
2220$~GeV, and $m_{V^\prime}<2310$~GeV are excluded.

%-------------------------------------------------------------------------------
%\section*{Acknowledgements}
%-------------------------------------------------------------------------------
% Not for Conf Notes according to Koji
% Acknowledgements for papers with collision data
% Version 15-Apr-2016

% Standard acknowledgements start here
%----------------------------------------------
We thank CERN for the very successful operation of the LHC, as well as the
support staff from our institutions without whom ATLAS could not be
operated efficiently.

We acknowledge the support of ANPCyT, Argentina; YerPhI, Armenia; ARC, Australia; BMWFW and FWF, Austria; ANAS, Azerbaijan; SSTC, Belarus; CNPq and FAPESP, Brazil; NSERC, NRC and CFI, Canada; CERN; CONICYT, Chile; CAS, MOST and NSFC, China; COLCIENCIAS, Colombia; MSMT CR, MPO CR and VSC CR, Czech Republic; DNRF and DNSRC, Denmark; IN2P3-CNRS, CEA-DSM/IRFU, France; GNSF, Georgia; BMBF, HGF, and MPG, Germany; GSRT, Greece; RGC, Hong Kong SAR, China; ISF, I-CORE and Benoziyo Center, Israel; INFN, Italy; MEXT and JSPS, Japan; CNRST, Morocco; FOM and NWO, Netherlands; RCN, Norway; MNiSW and NCN, Poland; FCT, Portugal; MNE/IFA, Romania; MES of Russia and NRC KI, Russian Federation; JINR; MESTD, Serbia; MSSR, Slovakia; ARRS and MIZ\v{S}, Slovenia; DST/NRF, South Africa; MINECO, Spain; SRC and Wallenberg Foundation, Sweden; SERI, SNSF and Cantons of Bern and Geneva, Switzerland; MOST, Taiwan; TAEK, Turkey; STFC, United Kingdom; DOE and NSF, United States of America. In addition, individual groups and members have received support from BCKDF, the Canada Council, CANARIE, CRC, Compute Canada, FQRNT, and the Ontario Innovation Trust, Canada; EPLANET, ERC, FP7, Horizon 2020 and Marie Sk{\l}odowska-Curie Actions, European Union; Investissements d'Avenir Labex and Idex, ANR, R{\'e}gion Auvergne and Fondation Partager le Savoir, France; DFG and AvH Foundation, Germany; Herakleitos, Thales and Aristeia programmes co-financed by EU-ESF and the Greek NSRF; BSF, GIF and Minerva, Israel; BRF, Norway; Generalitat de Catalunya, Generalitat Valenciana, Spain; the Royal Society and Leverhulme Trust, United Kingdom.

The crucial computing support from all WLCG partners is acknowledged
gratefully, in particular from CERN and the ATLAS Tier-1 facilities at
TRIUMF (Canada), NDGF (Denmark, Norway, Sweden), CC-IN2P3 (France),
KIT/GridKA (Germany), INFN-CNAF (Italy), NL-T1 (Netherlands), PIC (Spain),
ASGC (Taiwan), RAL (UK) and BNL (USA) and in the Tier-2 facilities
worldwide.
%----------------------------------------------

%The \texttt{atlaslatex} package contains the acknowledgements that were valid 
%at the time of the release you are using.
%These can be found in the \texttt{acknowledgements} subdirectory.
%When your ATLAS paper or PUB/CONF note is ready to be published,
%download the latest set of acknowledgements from:\\
%\url{https://twiki.cern.ch/twiki/bin/view/AtlasProtected/PubComAcknowledgements}

%The supporting notes for the analysis should also contain a list of contributors.
%This information should usually be included in \texttt{mydocument-metadata.tex}.
%The list should be printed either here or before the table of contents.

%-------------------------------------------------------------------------------
%\clearpage
%\appendix
%\part*{Appendix}
%\addcontentsline{toc}{part}{Appendix}
%-------------------------------------------------------------------------------

%In a paper, an appendix is used for technical details that would otherwise disturb the flow of the paper.
%Such an appendix should be printed before the Bibliography.

%-------------------------------------------------------------------------------
% If you use biblatex and either biber or bibtex to process the bibliography
% just say \printbibliography here
\printbibliography
% If you want to use the traditional BibTeX you need to use the syntax below.
%\bibliographystyle{bibtex/bst/atlasBibStyleWoTitle}
%\bibliography{VHConf,bibtex/bib/ATLAS}
%-------------------------------------------------------------------------------

\newpage
% ATLAS Collaboration author list
% Data extracted on 06-Jun-2016 for paper reference EXOT-2015-18
\begin{flushleft}
{\Large The ATLAS Collaboration}

\bigskip

M.~Aaboud$^\textrm{\scriptsize 135d}$,
G.~Aad$^\textrm{\scriptsize 86}$,
B.~Abbott$^\textrm{\scriptsize 113}$,
J.~Abdallah$^\textrm{\scriptsize 64}$,
O.~Abdinov$^\textrm{\scriptsize 12}$,
B.~Abeloos$^\textrm{\scriptsize 117}$,
R.~Aben$^\textrm{\scriptsize 107}$,
O.S.~AbouZeid$^\textrm{\scriptsize 137}$,
N.L.~Abraham$^\textrm{\scriptsize 149}$,
H.~Abramowicz$^\textrm{\scriptsize 153}$
% ...
\bigskip
\\
$^{1}$ Department of Physics, University of Adelaide, Adelaide, Australia\\
% ...
$^{12}$ Institute of Physics, Azerbaijan Academy of Sciences, Baku, Azerbaijan\\
% ...
$^{64}$ University of Iowa, Iowa City IA, United States of America\\
% ...
$^{86}$ CPPM, Aix-Marseille Universit{\'e} and CNRS/IN2P3, Marseille, France\\
% ...
$^{107}$ Nikhef National Institute for Subatomic Physics and University of Amsterdam, Amsterdam, Netherlands\\
% ...
$^{113}$ Homer L. Dodge Department of Physics and Astronomy, University of Oklahoma, Norman OK, United States of America\\
% ...
$^{117}$ LAL, Univ. Paris-Sud, CNRS/IN2P3, Universit{\'e} Paris-Saclay, Orsay, France\\
% ...
$^{135}$ $^{(a)}$ Facult{\'e} des Sciences Ain Chock, R{\'e}seau Universitaire de Physique des Hautes Energies - Universit{\'e} Hassan II, Casablanca; $^{(b)}$ Centre National de l'Energie des Sciences Techniques Nucleaires, Rabat; $^{(c)}$ Facult{\'e} des Sciences Semlalia, Universit{\'e} Cadi Ayyad, LPHEA-Marrakech; $^{(d)}$ Facult{\'e} des Sciences, Universit{\'e} Mohamed Premier and LPTPM, Oujda; $^{(e)}$ Facult{\'e} des sciences, Universit{\'e} Mohammed V, Rabat, Morocco\\
% ...
$^{137}$ Santa Cruz Institute for Particle Physics, University of California Santa Cruz, Santa Cruz CA, United States of America\\
% ...
$^{149}$ Department of Physics and Astronomy, University of Sussex, Brighton, United Kingdom\\
% ...
$^{153}$ Raymond and Beverly Sackler School of Physics and Astronomy, Tel Aviv University, Tel Aviv, Israel\\
% ...
$^{a}$ Also at Department of Physics, King's College London, London, United Kingdom\\
$^{b}$ Also at Institute of Physics, Azerbaijan Academy of Sciences, Baku, Azerbaijan\\
$^{c}$ Also at Novosibirsk State University, Novosibirsk, Russia\\
$^{d}$ Also at TRIUMF, Vancouver BC, Canada\\
$^{e}$ Also at Department of Physics {\&} Astronomy, University of Louisville, Louisville, KY, United States of America\\
% ...
\end{flushleft}

% Created with xml2latex.py

\end{document}